# One and Two-individual Movements of Fish after Chemical Exposure

**Quang Kha Quach and Tae-Soo Chon***

***Department of Biological Sciences, Pusan National University, Busan 609-735***

**Hungsoo Kim**

***Department of Biological Sciences, Pusan National University, Busan 609-735, and***

***SPENALO National Robotics Research Center, Pusan National University, Busan 609-735***

**Tuyen Van Nguyen**

***Department of Biological Sciences, Pusan National University, Busan 609-735, and***

***Department of Mathematics, Pusan National University, Busan 609-735***

Movement behavior of an indicator species, zebrafish (*Danio rerio*), was analyzed with one- and two-individual groups before and after treatment with a toxic chemical, formaldehyde, at a low concentration (1 ppm). After the boundary area had been determined based on experimental data, intermittency was defined as the probability distributions of the shadowing time during which data were above a pre-determined threshold and were obtained from experimental time-series data on the forces and the inter-distances for one and two individuals. Overall intermittencies were similar in the boundary and central areas. However, the intermittencies were remarkably different between the one- and the two-individual groups: the single line was used to fit the data for the one-individual group whereas two phases were observed with breakpoints (approximately 10 seconds in logarithm) in the exponential fitting curves for the two-individual group. A difference in the probability distributions of the shadowing time was observed "before" and "after" treatment for different areas. Intermittency patterns before and after treatment were contrasted in the center for the one-individual group whereas the difference was observed in the boundary for two-individual group. The intermittencies for the inter-distances of two individuals in the boundary and the central areas were markedly different before and after treatment. When the differences between the intermittencies in the boundary and the central areas and between "before" and "after" treatment are considered, the distribution patterns of the shadowing time (scaling behaviors or intermittency patterns) should be a useful means of bio-monitoring to detect contaminants in the environment.

*****Email: tschon@pusan.ac.kr;

Fax: +82-51-512-2262

## I. INTRODUCTION

The analysis of the response behaviors of animals has received considerable attention regarding *in situ* monitoring of indicator species since computational methods and interfacing techniques were introduced in the 1980's [1-4]. Monitoring by using behavioral changes is ecologically relevant, economical and faster than monitoring by using method of chemical detection [5-7]. Due to the high degree of complexity in behavioral data, however, various computational methods have been proposed to exploring time-series data on animal movements [1, 8]: parameterization with a fractal dimension [9] and permutation entropy [10, 11], statistical methods using correlation analyses [10, 12, 13], data transform including Fourier transforms [7, 14] and wavelet analysis [15]. Considering the complexity of behavioral data, informatics has been further applied to movement patterns, including self-organizing map [7, 14, 16] and multi-layer perception [6, 15], and is capable of identifying specific response behaviors of indicator species under chemical stress. Because of the uncertainties in behavioral patterns, the hidden Markov model has been used to analyze behavioral state changes after exposure to chemical treatment [16, 17]. However, the abovementioned reports mostly focused on data for single individuals, and not many studies were conducted on the responses of multiple individuals.

Regarding group formation by multiple individuals, simulation models based on the equations of motion have been proposed to elucidate the collective behavior associated with self-propelled particle systems according to the group (i.e., overall average orientation) and the neighbor (e.g., attraction, repulsion) responses [18-23]. Group behavior models were also analyzed, and observed data were evaluated; force components of individuals in collective motion were calculated in order to explain the relationship between the individual itself, its neighbors and environmental factors [24, 25]; individual fish movements were expressed by using the mass, drag coefficient, and external forces. Recently, the importance of nearest-neighbor interactions in group formation was addressed [26, 27].

In this study, we focused on the physical forces produced by one and two individuals under stressful conditions due to chemical exposure. In order to reveal the structure property in the movement data, we addressed the probability distributions of the shadowing time in time-series force data on fish observed in a confined area. Scaling behavior has been increasingly used in analyzing movement behavioral patterns of animals in the wild and in the laboratory. Intermittency is defined as the probability distribution of the shadowing time during which the data are consecutively higher than a threshold number [28-31]. For time-series data generated from a chaotic system (e.g., attractor), intermittency exhibits a universal algebraic scaling at high frequencies with a slope approximately $-3/2$ while it exhibits an exponential scaling at lower frequencies [28, 30].

Intermittency is among the universal mechanisms that produce chaos from a periodic orbit in a continuous way [32] and has been reported in various fields, including coordination of muscular systems [33, 34], chemical kinetics [35, 36], laser models [37], and fluid dynamics [38]. In ecology, flow intermittency regarding biodiversity determination in stream ecosystems has been recently investigated [39, 40]. Intermittency has been further applied to behavior studies. Harnos *et al*. [41] analyzed scaling and intermittency in the temporal behavior of nesting gilts, reporting that the time spent by a gilt in a given form of activity had a power-law probability distribution, and showed the intermittent occurrence of certain periodic behavioral sequences to indicate a critical state. Mashnova *et al*. [42] investigated intermittency and a truncated power law in aphid movement and addressed the alternate appearance of fast and slow movement phases that were distinguished by a threshold value of velocity. However, intermittency in response behavior of animals under chemical stress has not been extensively studied.

In addition to chemical response, we further observed individual movement at different locations in a confined area. Although test animals can move around in a more-or-less straightforward manner over a wide range, the individuals are constrained inside a confined arena within a boundary, especially for behavior monitoring within an observation arena [16]. The boundary zone was considered to be the area in which free movement would be minimally allowed, and is important for the life events, including protection and exploitation, of animals [43, 44]. We showed that the scaling behaviors of two individuals of *D. rerio* would be different at the boundary and the central areas of the observation arena before and after chemical treatment. Specifically, we intended to characterize intermittency in response behaviors in three different categories, 1) comparison of one and two individuals, 2) boundary and central areas, and 3) before and after chemical treatment. We analyzed the probability distributions of the shadowing time to address changes in the structure property in the movement data and found that intermittency in individual and group movement could be used as a possible means of behavioral monitoring.

# II. EXPERIMENTS

## 1. Test Organisms

One and two individuals of zebrafish, *D. rerio,* were observed under chemical stress. Due to vulnerability to chemical stress and availability of biological information (e.g., genomics, physiological responses), the zebrafish is considered to be one of the most suitable vertebrate model organisms for various biological tests [45-47], including behavior assessment [7, 16, 48]. The species has a strong potential for being an indicator in risk assessment [16, 49]. Individuals of wild-type *D. rerio* were obtained from a local fish dealer for stock population (300 individuals) and were reared for 2 weeks before observation [50] at a temperature of 25 ± 1$^{o}$C and pH of 7.1 ± 0.3 under a light/dark cycle of 14/10 h, light on at 7:00 h and off at 20:00 h [51]. Two fluorescent lights (26 J/s) were placed 50 cm above the rearing container. Tap water was filtered with air stones under air compression (DT - 10F, Chuang Xing Electric Appliances®) after dechlorination for three days. Fishes were fed dry food (Nutron Hi – Fi, PRODAC®) twice a day (once a day on weekends). Other rearing conditions are described below [16].

Test organisms (ages: 5 – 6 months; body lengths: 30 – 40 mm) were randomly chosen from the stock population and were placed individually in a glass aquarium (300 mm × 300 mm × 300 mm; water height of 20 mm). Before observation, organisms were acclimated to the observation system for 30 minutes [50]. To simplify observation and minimize noise, food and oxygen were not supplied to the arena during the observation period. Two 13J/s fluorescent lights were provided 50 cm above the water's surface and the two light sources were symmetrically 32 cm away from the center over the observation arena. Other rearing and observation conditions were the same as those used to rear the stock population.

Formaldehyde (HCHO, 37wt. % solution in water, A.C.S. reagent, Gamma–Aldrich®) was used as a source of stress to the test organisms. Formaldehyde is claimed to be one of most toxic environmental hormones and a possible carcinogenic agent through bioaccumulation [52]. The chemical was directly added to the water in the observation aquarium at a concentration of 1 ppm. In order to minimize noise, the chemical was delivered through an injector (Pipetman® P20) connected to the observation aquarium through a flexible polyethylene tube (1.85 mm in diameter and 1 m in length) after dilution with a proper amount of water.

## 2. Observations and Recording

The observation system consisted of an observation aquarium, a camera (Logitech®Vid™HD), a PC (Intel® Core ™ 2 Duo CPU E4500@ 2.20GHz ), and software for tracking the motion of multiple individuals. The software was developed in the Ecosystem and Behavior Lab. at Pusan National University based on stereo vision [53] after evaluation with a multiple individual tracking program (SynthEyes, 2008, Anderson Technologies LLC). The *x*-*y* position of each individual was continuously recorded at 30 frames per second from a top view in two dimensions before (30 minutes) and after (30 minutes) treatment. Five-minute segments were selected for analysis according to Suzuki *et al.* [27] and Herbert-Read *et al*. [54]. After treatment, fish immediately responded to olfactory stimulus from the chemical for approximately 5 minutes, showing abnormal behaviors including shaking and turning. Afterwards abnormal behaviors occurred less frequently. Movement tracks for the initial five minutes were analyzed before and after treatment.

Based on preliminary research [16, 55], a time segment of 0.2 s for recording movement was selected for this study. Because we aimed to observe overall movement changes of the fish specimens in two dimensions in response to the chemical treatment, the 0.2 s segment was sufficiently short for presenting the displacement of organism location [7, 14]. Extremely short-time response behaviors due to intoxication (e.g., compulsion, trembling) may be expressed in timer shorter than 0.2 s, but this type of behaviors of extremely short duration was not analyzed in this study [16]. Each movement segment was determined with three points with two consecutive 0.1 s segment (0.2 s in total). The observation was repeated 20 times for each group of one and two individuals. Mean values of the linear and the angular speeds were obtained from movement segments for each individual during the observation period; subsequently, the mean values were calculated from the averages of all individuals (i.e., $n$ = 20, and 40 for the one- and the two-individual groups, respectively) before and after treatment.

## 3. Computational Methods

### 3.1. Determination of boundary and central areas

Although test animals can move in a more-or-less straightforward manner (i.e., free run length), the individuals are also located inside a confined arena for monitoring the observation arena, as stated above [16]. We defined the boundary and the central areas by measuring the velocity of single individuals. Figures 1(a) and (b) show the distributions of the $x$- and the $y$-component of the velocity along each coordinate before treatment, respectively. In order to determine the boundary area, we inspected the cumulative sum of the velocity data. In Figures 1(c) and (d), the cumulative sums of the two components of the velocity from the sides of the arena are presented.

Subsequently, the cumulative sum were fitted with the exponential function $A(1-e^{-\alpha x})$, here $A$ is a proper amplitude and $\alpha$ is the damping parameter, which is taken as the inverse of the width of boundary. By fitting the data with the exponential function, the boundary width in $x$-coordinates was evaluated as 19.23 mm and 10.53 mm at the left and the right sides of the $x$-coordinate and 20.00 mm and 10.53 mm at the left and the right sides of the $y$-coordinate. From the evaluated value of the width of boundary area from the edge of each side, the largest value, 20.00 mm, was chosen to define the boundary area. The obtained value was comparable to the boundary areas empirically based on the fish size [16].

### 3.2. Real forces of each individual

Based on our empirical data, we measured changes in the forces on the test individuals before and after treatment. Following the framework of classical mechanics, we defined the total force $\vec{f}_i$ on the $i^{\text{th}}$ focal fish as the sum of two real forces, the frictional force $\vec{f}_i^{\,fric}$ and self-driven force $\vec{f}_i^{\,d}$ :

$$\vec{f}_i = \vec{f}_i^{\,d} + \vec{f}_i^{\,fric}, \qquad (1)$$

with

$$\vec{f}_i = m\vec{a}_i \qquad (2)$$

$$\vec{f}_i^{\,fric} = \mu\vec{v}_i, \qquad (3)$$

where $m$ is the mass of the fish, and $\mu$ is the friction coefficient in water. However, in our analysis, the mass $m$ is set to unity, and $\mu$ is assumed to be 0.05 [21].

To calculate the self-driven force $\vec{f}_i^{\,d}$, we calculated the velocity and the acceleration of $i^{\text{th}}$ individual at time $t$ by using $\vec{v}_i = \frac{\Delta\vec{r}_i}{\Delta t}$, $\vec{a}_i = \frac{\Delta\vec{v}_i}{\Delta t}$ from the movement tracks. We directly calculated the $x$- and the $y$-components and the absolute force for the one-individual group while forces were calculated according to center of mass, individual forces, and the relative coordinate between two individuals in the two-individual group.

### 3.3. Calculation of intermittency

The mean value of the absolute of the force measured before treatment was used as a criterion to determine the threshold for the shadowing time (Figure 2). We used one fourth of the mean value as the threshold, after testing various levels of the threshold from one eighth to 2 times the mean value. One fourth the mean value was most suitable in characterizing the probability distributions of the shadowing time in the boundary and the central areas, as well as “before” and “after” treatment. The threshold value based on the absolute value of force was also used for the $x$-component and the $y$-component.

The shadowing times and their probability distributions were expressed on a logarithmic scale. The slopes and the elevations were obtained using a regression analysis [56]. The probability distribution of shadowing times of long duration was also fitted to an exponential curve when breakpoints occurred in intermittency [28, 30].

## III. RESULTS

Figure 3 shows the probability distributions of the shadowing time for forces on individuals observed in the boundary and the central areas when the time duration was selected according to the threshold (95.51 mm/s$^2$) in one- and two-individual groups. For the *x*- and the *y*-components of the forces, the probability distributions of the shadowing time were overall similar between the boundary and the central areas, but were different between one- and two-individual groups (Figures 3(a) – (b), (d) – (e), (g) – (h), and (j) – (k)). Linearity across different shadowing times was observed for the one-individual group (Figures 3(a) – (f)) whereas the linearity was not sustained and probability patterns appeared in the curve for the two-individual group (Figures 3(g) – (l)). The slopes of the distribution became steeper for long-time duration (i.e., right-hand side of the *x*-axis) for the short-time duration. For the *x*- and the *y*-components, the probability distributions of the shadowing time in the boundary area (Figures 3(a) – (b)) appeared to be slightly steeper than that in the central area (Figures 3(d) and (e)), but no statistical difference was observed in the regression lines according to their slopes ($p>0.05$) [56].

For the absolute forces, although the probability distributions of shadowing time were, in general, similar to the *x*- and the *y*-components, a difference was observed to some degree in the shapes of the probability distributions (Figures 3(c), (f), (i), and (l)). Before treatment in the boundary area, for instance, the log abundance for the long-time duration appeared to spread over a broader range (i.e., long foot at the right bottom corner in Figure 3(c)) whereas this type of long foot was not observed in the central area. In the two-individual group, similarly, distribution patterns were different in the boundary and the central areas, as well as “before” and “after” treatment (Figures 3(i) and (l)). A detailed description of the distribution patterns, however, is beyond the scope of this study and will be reported elsewhere.

Intermittency was further contrasted before and after treatment in Figure 4, and the probability distributions of the shadowing time were fitted to lines and exponential curves because intermittency exhibited a universal algebraic scaling at high frequencies and an exponential scaling at lower frequencies [28, 30]. Table 1 summarizes the slopes of the lines based on regression analyses, and Table 2 lists the coefficients and fittings to the exponential functions in the boundary and the central areas for one- and two-individual groups. For the one-individual group, the probability distributions were fitted to single lines (Figures 4(a) and (b), (d) and (e)). The slopes were similar and were in the range of -1.89 – -1.91 for the *x*- component and -1.75 – -1.76 for the *y*-components before and after treatment, but the difference in the slopes of the regression lines were not statistically significant ($p>0.05$; Figures 4(a) and (b)) [56]. In the central area, however, the slopes were different for the *x*-and the *y*-components of the forces. The slopes were statistically steeper for both components of the forces after treatment (-1.79 – -1.80) than "before" treatment (-1.13 – -1.32) ($p<0.05$; Figures 4(d) and (e)), indicating that the phase change in the shadowing time was more sensitive in the central area under chemical stress.

For the absolute forces, the probability distributions of the shadowing time for one individual (Figures 4(c) and 4(f)) were more spread compared to those for the *x*-and the *y*-components of the forces in the boundary and the central areas. The slopes appeared to be different, with statistical significance, before and after treatment (Table 1). At the boundary area, slopes were steeper after treatment (-1.38) than before treatment (-1.02) whereas slopes were less steep in the central area after treatment (-1.34) than before treatment (-1.66) (Table 1).

Forces on the center of mass for the two-individual group were also calculated (Figures 4(g) – (l)). Compared with the forces on the one-individual group (Figures 4(a) – (f)), the probability distributions of the shadowing time were different, showing two phases as stated above. In both the *x*- and the *y*-components, intermittency appeared to be curved with a breakpoint in the boundary (Figures 4(g) – (h)) and the central (Figures 4(j) – (k)) areas whereas single lines were fitted to the intermittency curves for the case of the one-individual group (Figures 4(a), (b) and (d), (e)), as stated above. The breakpoint was found to be around 10 seconds, and intermittency was overall similar between the boundary and the central areas for the two-individual group. It was remarkable that the difference in intermittency before and after treatment was more clearly observed in the boundary area (Figures 4(g) and (h)), contrary to the case of the one-individual group where the difference was only observed in the central area (Figures 4(d) and (e)) (Table 1). It is also noteworthy that after treatment, the elevation of the intermittency (i.e., intercepts of regression lines) was lower in both the *x*- and the *y*-components in the boundary area (Figures 4(g) and (h)). A statistical difference between the boundary and the central areas was observed for the absolute forces (Figures 4(i) and (l), Table 1).

For the absolute forces on two individuals, curves were also formed in the boundary area, more strongly for "after" treatment (Figure 4(i)) although higher variation was observed in probabilities compared to the *x* and the *y*-components of the forces (Figures 4(g) and (h)). The breakpoint appeared to slightly move toward long time duration, a little over 10 seconds (Figure 4(i)). The lines fitted to the probability distributions at high frequency (i.e., before breakpoint) were statistically different before (-0.89) and after (-1.19) treatment in the boundary area (Table 1). In the central area, however, single lines were fitted to probability distributions across the shadowing time, and the slopes (-0.94 and -0.96) were not statistically different (Table 1). We also fitted the intermittency at lower frequency (i.e., a long shadowing time after breakpoint) to an exponential function [28, 30]. The coefficients were in the range of 0.30 – 0.34 and exponential curves before and after treatment were not statistically different when the goodness of fit between the two curves was tested according to the chi-square test [56] (Table 2). Although not presented in the figures, intermittency curves for velocities observed at the boundary and the central areas were similar to the case of forces (center of mass) both “before” and “after” treatment. However, the intermittency of velocities was weaker in expressing the difference between “before” and “after” treatment.

We also calculated the relative forces between two individuals in the two-individual group (Figures 5(a) – (f)). Similar to force of center of mass (Figures 4(g) – (l)), two phases were observed around the breakpoint of 10 seconds. The shapes of the probability distributions before and after treatment were different in the boundary area while the shapes were similar in the central area. Statistical significance was observed in the lines fitted to the intermittency in the *x*- and the *y*-components for the short-time duration before and after treatment (Table 1) (Figures 5(a) and (b)). The slopes ranged from -0.90 to -1.10 before treatment and from -1.09 to -1.63 after treatment for the *x*- and the *y*-components in the boundary and the central areas. The slopes were statistically different before and after treatment in the boundary area (Table 1). Although the slopes were not different in the central area, the elevations (i.e., *y*-intercepts of the regression lines) were statistically different before and after treatment (see Ref. 56 for the statistical significance of the elevation in a regression line).

For absolute forces, the probability distributions were also different before and after treatment (Figures 5(c) and (f)). A breakpoint was observed, and the point appeared to move more toward the long-time duration, approximately matching 25 seconds in the boundary area. The slopes (-0.76 – -0.89) of the regression lines for the absolute force were less steep compared to those of the *x*- and the *y*-components (-0.90 – -1.63) in the boundary area and were comparable to those of the absolute forces (-0.94 –-0.96) on the center of mass in the center area (Table 1). The coefficients ($\alpha$) of the exponential functions were also fitted to the probability distribution of the long shadowing time and ranged from 0.30 to 0. 35. The exponential curves before and after treatment were not statistically different when the goodness of fit between two the curves was tested according to the chi-square test [56].

We also checked the distribution pattern for forces for all individuals in the two-individual group (Figures 5(g) – (l)). Similar to the case of intermittency of the relative force, overall probability distributions were observed in two phases, single lines for short shadowing times and exponential curves for long shadowing times (Figures 5(g) – (h) and (j) – (k)). According to this figure, the probability distributions for the shadowing time tended to be slightly steeper in both the boundary and the central areas after treatment. Difference in the slopes and the elevations were observed before and after treatment for the *x*- and the *y*-components, as well as the absolute forces, and these differences were statistically significant (Table 1). In the absolute forces, however, breakpoints were not clearly observed in the central area (Figure 5(l)). Overall, the difference in intermittency appeared to be more clearly observed in the boundary area (Table 1). Similar to the case of the relative force, intermittency of individual forces was fitted to an exponential function with $\alpha$ values ranging from 0.30 to 0.35, and the exponential curves before and after treatment were not statistically different, similar to two cases above [56] (Table 2).

We also calculated the probability distributions of the shadowing time for two individuals' inter-distance. The difference was outstanding in the boundary area before and after treatment; the curve became rapidly steeper after the breakpoint (Figure 6(a)). Similar to the case of forces, the break point was formed around 10 seconds. In the center, however, single lines were fitted both "before" and "after" treatment. The slopes of the intermittency appeared to be flat, ranging from 0.22 to -0.26 in the boundary area. The slope after treatment, however, became steeper (-0.80) than the slope before treatment (-0.47) in the central area (Figure 6(b)). The slopes before and after treatment were statistically different for both the boundary and the central areas (Table 1, Figure 6). Exponential functions were fitted to the intermittency after treatment, with $\alpha$ = 0.26 ($R^2$=0.71) in the boundary area, according to chi-square-test goodness of fit [56] (Table 2).

## IV. DISCUSSIONS AND CONCLUSIONS

It was remarkable that the data structure was fundamentally different between single and two individuals. The breakpoints with two phases in intermittency were observed for short and long shadowing times in the two-individual group (Figures 3(g), (h), (j) and (k)) whereas single lines were presented in the one-individual group (Figures 3(a), (b), (d), and (e)). The linearity and the breakpoints were consistently observed both “before” and “after” treatment (Figures 3 – 5). This indicates that pairwise interaction between two individuals played a key role in determining movement data structure. Recently, the importance of the nearest-neighbor relationship in group behavior was reported. Herbert-Read *et al*. [54] demonstrated the importance of repulsion and response to a single nearest neighbor in fish group–behavior dynamics. Pairwise interactions are important in qualitatively capturing the correct spatial interactions in small groups of fish when compared with the observed data [57]. Our study indirectly supports the significance of two-individual interactions in group formation.

In addition, the intermittency patterns were substantially different “before” and “after” chemical exposure for different areas in the observation arena (Figures 4 – 6). The probability distributions of the shadowing time were different before and after treatment in the center area for the one-individual group whereas the difference was observed in the boundary area for the two-individual group. Regarding behavioral-state changes (i.e., transition probability of different movement patterns), no qualitative difference was observed between the boundary and the central areas [16]. Indeed, the overall patterns of intermittency were similar in the boundary and the central areas (Figure 3). However, response to chemical stress appeared differently according to the organism’s location in the arena. Especially, the inter-distances between two individuals were markedly different “before” and “after” treatment (Figure 6). This further indicates that pairwise interactions are strongly reflected in the spatial dynamics in the boundary area, suggesting emergence of new property in the movement data structure in responding to neighbors nearby edge areas under stressful conditions. To the best of the authors’ knowledge, this is the first report observing differences in the intermittency of forces on individuals in the boundary and the central areas. Further study, however, is required in both the computational and the biological aspects.

The existence of breakpoints in two-individual groups (Figures 3 – 5) also reflects critical time duration for characterizing collective motion. Considering that the slopes for intermittency at short times were near -1.5 and the slopes became steeper, the time duration of 10 seconds may be due to behaviors stemming from the association of two individuals in a confined area (e.g., approach, communication). The breakpoints moved toward longer time duration in the case of absolute forces (Figures 5(c) and (i)). Currently, the mechanism of breakpoint formation is not known. This time duration may also be due to an output from physiological networks [58]. In biological aspects, physiological and/or molecular genetics networks could be investigated; how stereotypic changes in behavioral patterns could originate from integrative actions of neural and endocrine systems [50]. However, the detailed mechanism is currently unknown and more research may be required in this direction in the future.

Considering the difference in the intermittency patterns at different locations before and after treatment, especially in the boundary area, the probability distributions of the shadowing time could be utilized as a useful means of monitoring chemical stress. Intermittency in the inter-distance between two individuals was remarkably different between “before (i.e., strong curves with a breakpoint)” and “after (i.e., single line)” treatment in the boundary, as shown in Figure 5(a).

In this study, we did not use the abundance data for the minimal time duration (i.e., the first probability matching to the shortest shadowing time); the points were not maximal in all cases (Figures 3 – 6). For instance, the point matching minimum time duration in Figure 4(a) showed abundance less than the abundance shown by the second shortest shadowing time. Considering the negative value (-1.5) of the intermittency [30], the abundance should be theoretically maximized at the shortest shadowing time. The somewhat lower abundance at the minimum shadowing time indicates that the shortest shadowing time is expressed in a reserved manner biologically, and this may stem from the physiological and behavioral nature of the organisms. However, the reason is currently unknown. In some cases, sufficient data to evaluate intermittency were not recorded. For instance, the intermittency applied to the relative force in the central area, insufficient data points were collected in the central area (Figures 5(d) and (e)). This may be due to the fact that more data points were recorded in the boundary area. Considering that an acute response due to the olfactory stimulus of formaldehyde were generally observed within 5 minutes as stated above, the observation time may not be extended due to weaker response behaviors after 5 minutes, but the replication number may be increased. More data need to be accumulated in a future study.

In this study, only one concentration of the chemical was tested. More research is needed at different concentrations of chemicals in order to determine the fish's behavioral response to an increase in stress levels. In the future, more than two individuals could be tested, and the contributions of additional neighbors to group formation could be more closely investigated.

In conclusion, the intermittency of forces and inter-distances in one- and two-individual groups effectively addressed the structural changes in collective motion. Whereas linearity was observed in the probability distributions of the shadowing time for the one-individual group, two phases with breakpoints were measured for two-individual group consisting linearity (the short shadowing time) and exponential function (the long shadowing time). Furthermore, the effect of chemical stress was demonstrated by using difference between the intermittencies in the boundary and the central areas. Differences in the intermittency patterns appeared more clearly in the center for the one-individual group, but the differences were more effectively presented in the boundary for the two-individual group. Changes in the probability distributions of the shadowing time suggested that the pairwise association between two individuals is essential in collective motion and group formation. The sensitivities in the intermittencies evaluated for the one- and the two-individual groups in response to toxic chemicals can be utilized as a means of behavioral monitoring to detect contaminants in the environment.

**ACKNOWLEDGMENT**

This research was supported by the Basic Science Research Program through the National Research Foundation of Korea (NRF) funded by the Ministry of Education, Science and Technology (2011-0012960).

Table 1 Estimates of the slopes and elevations by applying a regression analysis to the intermittency of forces for movement of zebrafish in one- and two-individual groups in the boundary and central areas before and after chemical treatment.

| **No. of indi** | **Type** | **Treat** | **Boundary** | | | **Center** | | |
|---|---|---|---|---|---|---|---|---|
| | | | **X** | **Y** | **Absolute** | **X** | **Y** | **Absolute** |
| One | Indi | Before | -1.89±0.20 | -1.76±0.19 | -1.02±0.27 | -1.32±0.16 | -1.13±0.16 | -1.66±0.49 |
| | | After | -1.91±0.19 | -1.75±0.19 | -1.38±0.26* | -1.79±0.11* | -1.80±0.18* | -1.34±0.11* |
| Two | Center of mass | Before | -1.11±0.17 | -1.10±0.11 | -0.89±0.11 | -1.08±0.15 | -1.17±0.18 | -0.94±0.26 |
| | | After | -1.61±0.17* | -1.48 ±0.14* | -1.19±0.16* | -1.26±0.32* | -1.14±0.26* | -0.96±0.31 |
| | Relative | Before | -1.09 ±0.13 | -1.1±0.09 | -0.79±0.18 | -0.90±0.19 | -1.07±0.15 | -0.76±0.32 |
| | | After | -1.63±0.16* | -1.41±0.09* | -0.89±0.14① | -1.30±0.22② | -1.09±0.20③ | -0.87±0.40* |
| | Indi | Before | -1.35±0.11 | -1.28±0.08 | -0.71±0.15 | -0.94±0.10 | -1.07±0.12 | -0.79±0.06 |
| | | After | -1.61±0.17* | -1.52±0.09* | -0.89±0.16* | -1.03±0.19④ | -1.02±0.17⑤ | -1.03±0.07* |

* Indicates statistical significance "before" and "after" treatment based on the different slopes of the regression lines ($p<0.05$) [56].

Numbers in circles present statistical significances "before" and "after" treatment based on the different elevations in the regression lines ($p<0.05$) [56] ①-2.07/-1.50, ②-1.15/-0.97, ③-1.08/-0.81, ④-1.40/-1.17 and ⑤-1.47/-1.10, before /after treatment, respectively.

Table 2 Estimates of the coefficient ($\alpha$) of the exponential function and the goodness of fit (chi-square test) applied to the intermittency of forces of zebrafish in one- and two-individual groups before and after treatment.

| **Type** | **Treat Test** | **Boundary** | | **Center** | |
|---|---|---|---|---|---|
| | | **X** | **Y** | **X** | **Y** |
| Center of mass | Before † | 0.31 (0.42) | 0.30 (0.41) | 0.30 (0.59) | 0.30 (0.36) |
| | After † | 0.35 (0.61) | 0.34 (0.62) | 0.30 (0.30) | 0.30 (0.25) |
| | $\chi^2$ †† | 10.66 (0.15) | 10.14 (0.18) | 0.63 (1.00) | 1.79 (0.97) |
| Relative | Before | 0.30 (0.64) | 0.31 (0.45) | 0.30 (0.56) | 0.31 (0.27) |
| | After | 0.35 (0.79) | 0.33 (0.85) | 0.31 (0.30) | 0.31 (0.56) |
| | $\chi^2$ | 10.28 (0.17) | 11.72 (0.11) | 5.61 (0.59) | 11.89 (0.10) |
| Indi | Before | 0.33 (0.48) | 0.33 (0.52) | 0.30 (0.34) | 0.30 (0.29) |
| | After | 0.35 (0.58) | 0.35 (0.52) | 0.30 (0.61) | 0.31 (0.41) |
| | $\chi^2$ | 3.79 (0.80) | 3.19 (0.87) | 5.30 (0.63) | 7.67 (0.47) |

† Numbers in parentheses indicate the $R^2$ value according to the coefficient estimate of the functions (exponential decay [56]).

†† Numbers in parentheses present the probability according to chi-square test's goodness of fit between two exponential functions, one before and one after treatment [56].

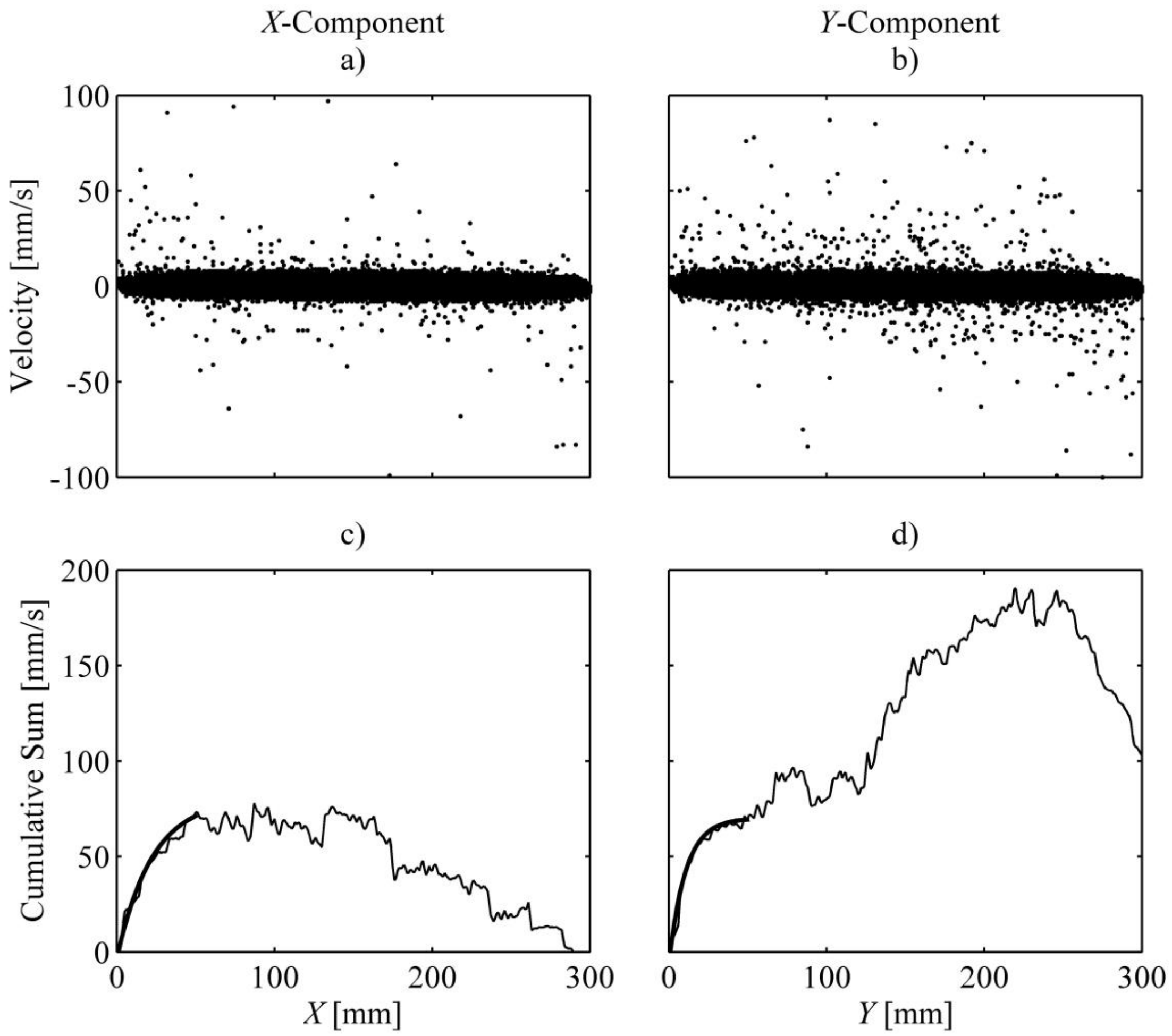

Fig. 1. Velocity distribution and its cumulative sum along each axis of fish movement in the observation arena in defining the boundary area: (a) *x*-component, (b) *y*-component, (c) cumulative sum of the *x*-component, and (d) cumulative sum of the *y*-component. Solid curves at the boundary (c) and (d) indicate the exponential curves fitting the data from 0 mm to 50 mm.

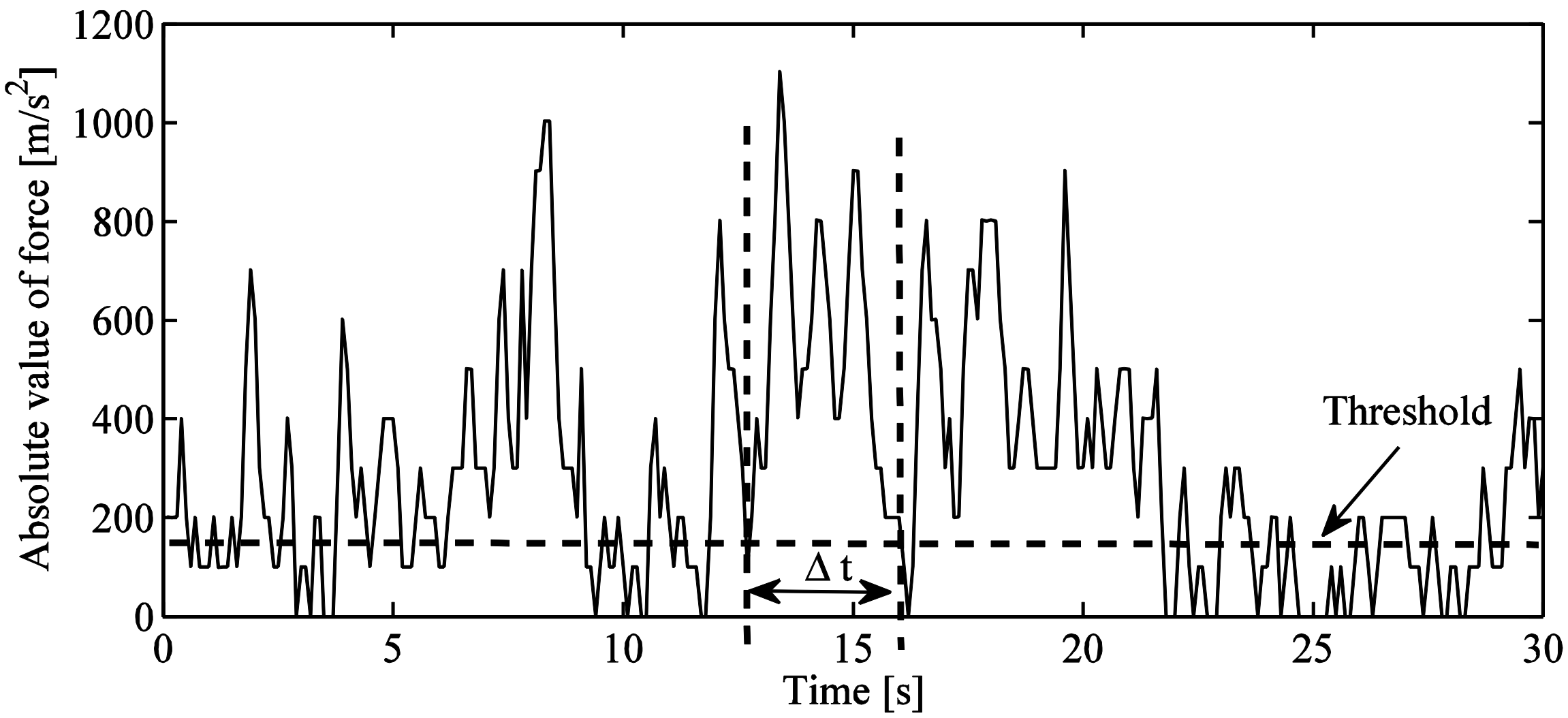

Fig. 2. Time series of the absolute value of force for one individual before treatment in the center for various shadowing times ($\Delta t$) and a threshold (dashed line) to determine the shadowing time.

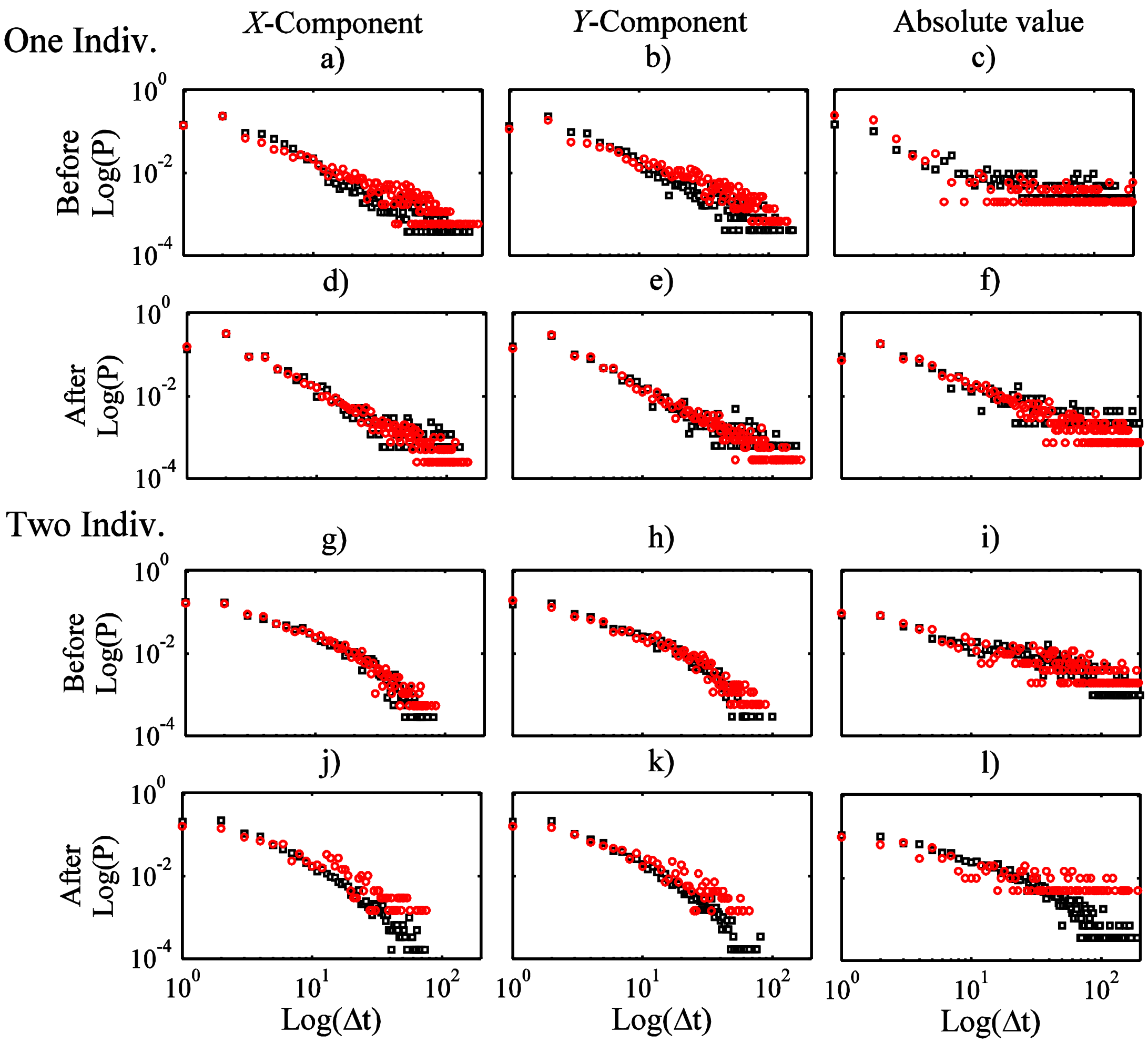

Fig. 3. Intermittency patterns for forces on one individual before (a, b, and c) and after (d, e, and f) treatment, and those for two individuals before (g, h, and i) and after (j, k, and l) treatment in the boundary (blank squares) and the central (red circles) areas. The probability distributions of the shadowing time were fitted to single lines for the one-individual group whereas they were matched to exponential curves for the two-individual group.

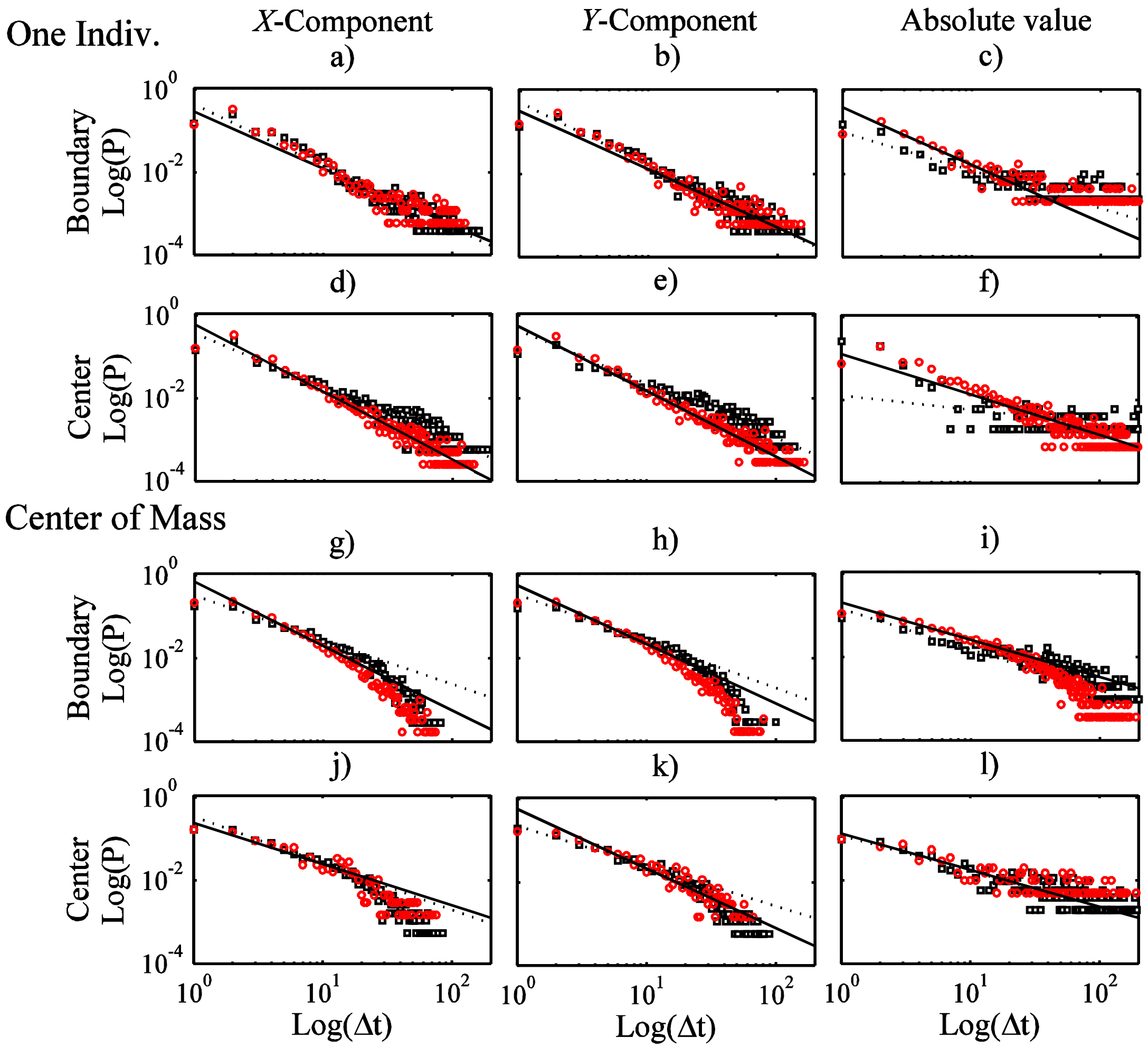

Fig. 4. Intermittency patterns for the forces on the one-individual group in the boundary area ((a) *x*-components, (b) *y*-component, and (c) absolute value) and in the central area ((d) *x*-component, (e) *y*-component, and (f) absolute value), and those for force at the center of mass of the two-individual group in the boundary area ((g) *x*-component, (h) *y*-component, and (i) absolute value) and the central area ((j) *x*-component, (k) *y*-component, and (l) absolute value). Intermittency patterns before and after treatment were different in the center for the one-individual group whereas the difference was observed in the boundary for the two-individual group. Solid and dotted lines fitting "before" and "after" treatment, respectively.

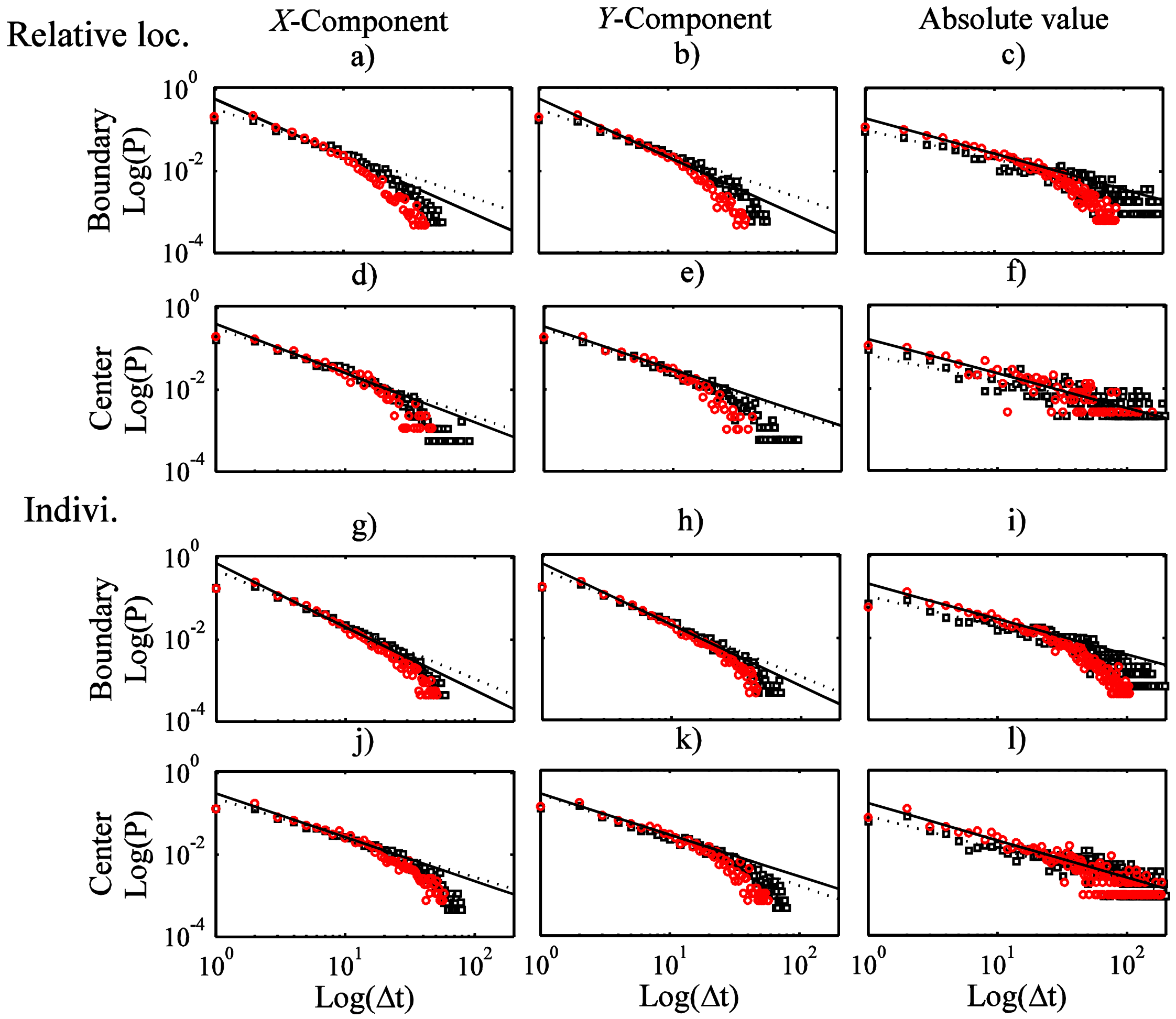

Fig. 5. Intermittency patterns for the forces before (blank squares) and after (red circles) treatment in the two- individual group. The relative force on individuals in the boundary area ((a) *x*-component, (b) *y*-component, and (c) absolute value) and in the central area ((d) *x*-component, (e) *y*-component, and (f) absolute value), and those on two individuals in the boundary area ((g) *x*-component, (h) *y*-component, and (i) absolute value) and in the central area ((j) *x*-component, (k) *y*-component, and (l) absolute value). Differences in the intermittency patterns before and after treatment were more clearly observed in the boundary for relative forces whereas the difference was equally observed in the boundary and the center for individual forces. Solid and dotted lines fitting "before" and "after" treatment, respectively.

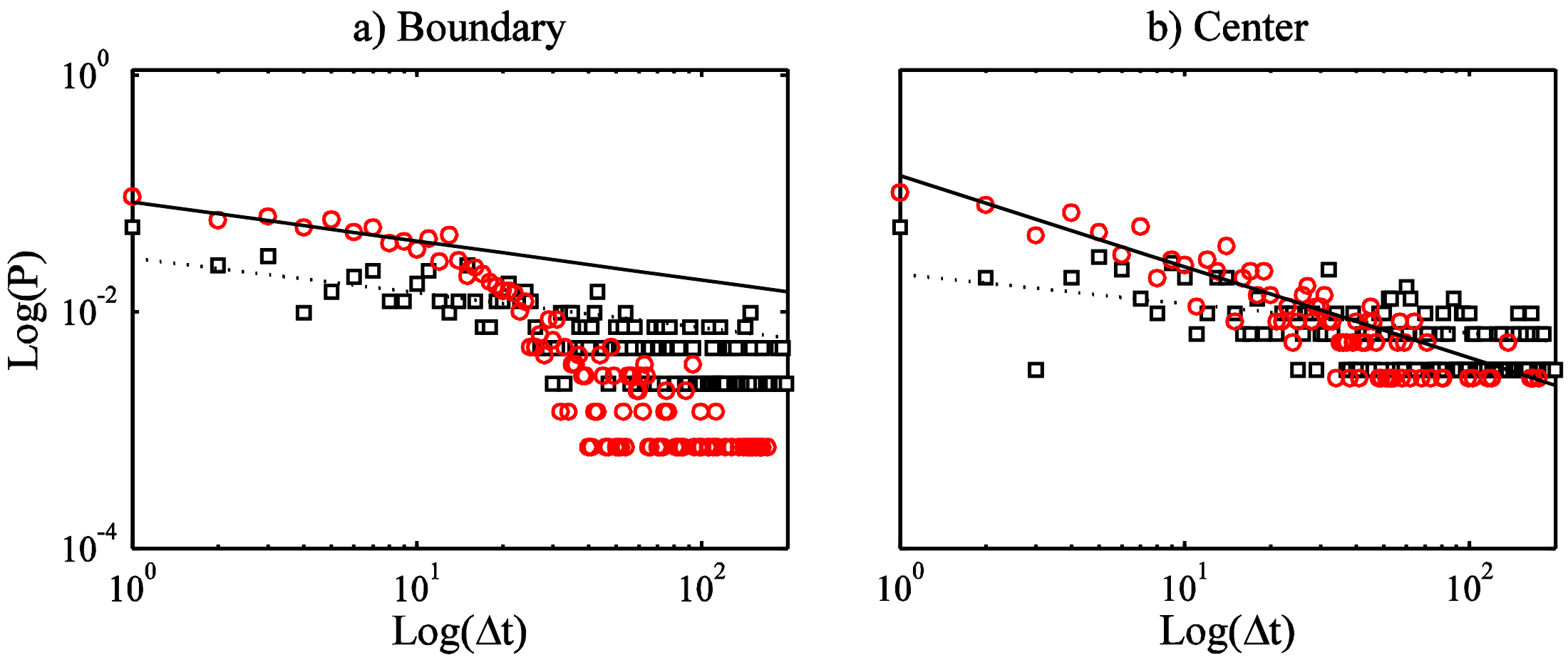

Fig. 6. Intermittency patterns for the inter-distance between two individuals before (blank squares) and after (red circles) treatment in (a) the boundary (slopes before (-0.22 ± 0.08) and after (-0.26 ± 0.12) treatment), and (b) the center (slopes before (-0.47 ± 0.06) and after (-0.80 ± 0.11) treatment). The intermittency pattern was markedly different after treatment in the boundary with a breakpoint clearly separating flat (for short shadowing time) and steep (for long shadowing time) slopes. Solid and dotted lines fitting "before" and "after" treatment, respectively.